\documentclass[aps,prb,manuscript]{revtex4}
\usepackage{graphicx}
\usepackage{epsfig}
\usepackage{color}
\usepackage{dcolumn}
\usepackage{bm}

\begin{document}

\title{Effects of site asymmetry and valley mixing on Hofstadter-type spectra of bilayer graphene in a square-scatter array potential}
\author{Danhong Huang$^{1,2}\footnote{Corresponding author's email: danhong.huang@us.af.mil}$, Andrii Iurov$^{3}$, Godfrey Gumbs$^{4,5}$ and Liubov Zhemchuzhna$^{3}$}
\affiliation{
$^{1}$Air Force Research Laboratory, Space Vehicles Directorate, Kirtland Air Force Base, NM 87117, USA\\
$^{2}$Department of Electrical \& Computer Engineering, University of New Mexico, Albuquerque, NM 87131, USA\\
$^{3}$Center for High Technology Materials, University of New Mexico, Albuquerque, NM 87106, USA\\
$^{4}$Department of Physics and Astronomy, Hunter College of the City University of New York, 695 Park Avenue, New York, NY 10065, USA\\
$^{5}$Donostia International Physics Center (DIPC), P de Manuel Lardizabal, 4, 20018, San Sebastian, Basque Country, Spain}
\date{\today}

\begin{abstract}
Under a magnetic field perpendicular to an monolayer graphene, the existence of a two-dimensional periodic scatter array can not only mix Landau
levels of the same valley for displaying split electron-hole Hofstadter-type energy spectra, but also couple two
sets of Landau subbands from different valleys in a bilayer graphene. Such a valley mixing effect with a strong scattering strength has been found observable and studied thoroughly in
this paper by using a Bloch-wave expansion approach and a projected $2\times 2$ effective Hamiltonian including interlayer effective mass, interlayer coupling and asymmetrical
on-site energies due to a vertically-applied electric field.
For bilayer graphene, we find two important characteristics, i.e., mixing and interference of intervalley scatterings in the presence of a scatter array,
as well as a perpendicular-field induced site-energy asymmetry which deforms severely or even destroy completely the Hofstadter-type band structures due to the dependence of
Bloch-wave expansion coefficients on the applied electric field.
\end{abstract}
\pacs{PACS:}
\maketitle

\section{Introduction}
\label{s1}

Shortly after its discovery and fabrication in 2004, graphene has captured tremendous attention and generated an enormous wave of research
activities due to its unique Dirac-cone-type electronic band-structures and properties.\,\cite{gr1,gr2,gr3} This wave also includes a huge amount of research works concerned with magnetic-field behavior, 
electronic properties, Landau levels (LLs)\,\cite{grm1,grm2} and quantum Hall effect\,\cite{grm11,grm12,grm13}.
Nearly at the same time, bilayer graphene (BLG), which consists of two closely-located graphene sheets, was also 
fabricated and tested experimentally.\,\cite{blg01,blg02,blg03}  The BLG electronic properties are found significantly different depending on details of an A-B stacking process, or Bernal-stacked form, with relatively shifted carbon-atom positions
in two layers.\,\cite{blg11} Bilayer graphene revealed some highly unusual properties, e.g., unconventional quantum Hall
effect\,\cite{bil2e} and cyclotron resonance\,\cite{bil3}.
\medskip

A comprehensive theoretical study of the LL degeneracy and quantum Hall effect for BLG
in Bernal stacking was reported in Ref.\,[\onlinecite{BilM1}]. Based on an effective two-dimensional Hamiltonian, it was concluded that
the low-energy spectrum of BLG can be characterized as parabolic dispersion of chiral quasi-particles with a Barry phase $2\pi$. Meanwhile, its magnetic-filed dependent energy spectrum is found
consisting of a set of nearly equidistant four-fold degenerateLLs. In this paper, we will employ such an effective-Hamiltonian approach to establish theoretical formalism for modulated
LLs in the presence of a square-scatter array potential in Sec.\,\ref{s2}.
\medskip

One of the most unusual and fascinating phenomena related to the electronic spectrum under a perpendicular quantizing magnetic field is the
so-called {\em Hofstadter butterfly}\,\cite{hofs-main, Azbel}, theoretically predicted in 1976. Here, a recursive fractal electron spectrum
was obtained as a function of prime ratio of the magnetic flux passing through a lattice unit cell to a fundamental flux quanta,
and these degenerate electronic subbands split and clustered themselves into different patterns corresponding to the value of a given magnetic-flux ratio.
By performing first-principles calculations for hexagonal two-dimensional graphene-type lattice, tight-binding approximation resulted in a
Hofstadter-butterfly-like clustering pattern, except for an asymmetry with respect to zero wave vector\,\cite{paula}.
Such types of Hofstadter band-structure were also predicted to exist in carbon nanotubes as pseudo-fractal
magneto-electronic spectrum\,\cite{nemec2} and also in bilayer graphene\,\cite{nemec1}.
\medskip

In a recent experiment, Hofstadter's butterfly and fractal quantum Hall effect have been extended to Moire
superlattices, which are formed as BLG or flakes are coupled to a rotationally aligned hexagonal
boron nitride layer\,\cite{dean2013,ponom,woods1} inside a van der Waals
heterostructure sample.\,\cite{huntM} The main idea involved in such experiments is that an elementary lattice-unit cell through which the magnetic
flux was measured\,\cite{hofs-main} will be replaced by a much bigger supercell of the Moire lattice,\,\cite{schmidt,yang,wang1} so that the
butterfly is expected to be seen at a much lower magnetic field. Additionally, the theory for such butterfly structures in twisted BLG
was proposed in Ref.\,[\onlinecite{bis}], in which long-period spatial patterns can be created precisely at small
twist angles. Later, the coexistence of both fractional-quantum-Hall and integral-quantum-Hall states associated with 
fractal Hofstadter spectrum was confirmed experimentally 
within such twisted-bilayer structures.\,\cite{wangZ} Moreover,
specific subband gaps of a Hofstadter's butterfly were also found for interacting Dirac fermions in graphene.\,\cite{apalkov1}
\medskip

On the other hand, in the absence of a magnetic field, a periodic electrostatic field gives rise to new zero-energy states
with minigaps and chirality\,\cite{brey}, and their composite wave functions can still satisfy the required Bloch periodic condition. Apart from
this, new massless Dirac fermions with strong anisotropic properties\,\cite{park2} are realized in graphene subjected to a slowly-varying
periodic potential.\,\cite{park1} In contrast, a spatially-uniform interaction of Dirac electron with an off-resonant optical field can lead to the formation of either
gapped\,\cite{kibis1,our11,our12} or anisotropic dressed\,\cite{kibis2} states depending on polarizations of an imposed irradiation.
\medskip

Very interestingly, two unique features associated with BLG system have been found. The first property is the intervalley mixing and the quantum interference effect coming from two valleys in the presence of a two-dimensional scattering-lattice potential,
while the second property results from a site-energy asymmetry induced by a perpendicular electric field. Here, the latter factor is able to destroy the Hofstadter-type fractal band structures established by an in-plane scattering-lattice potential 
and an out-of-plane quantizing magnetic field, resulting in strongly deformed self-repeated patterns. Such a phenomena is attributed to the dependence of Bloch-wave expansion coefficients on an applied electric field, leading to an electro-modulation of the Hofstadter-type subband splittings.
\medskip

The rest of the paper is organized as follows. In Sec.\,\ref{s2} we present theoretical formalism and acquire a set of characteristic equations,
describing electron energy spectrum and corresponding eigenstates for BLG in the presence of both a perpendicular quantizing magnetic field
and a two-dimensional periodic electrostatic modulation potential. These results expand the previously studies for a two-dimensional electron gas\,\cite{kuhn}
and for a monolayer graphene\,\cite{our1,ourRK}. In Sec.\,\ref{s3}, we display and discuss our numerical results demonstrating fractal Hofstadter band-structures
in different ranges of magnetic field of interest and with various modulation strengths in a close up view for separate LLs and self-repeated superstructures as well. 
Finally, a brief summary with remarks is given in Sec.\,\ref{s4}.

\section{Model and Theory}
\label{s2}

By considering $K$ and $\tilde{K}$ valleys, where $\displaystyle{K=(\frac{2\pi}{3a},\,\frac{2\pi}{\sqrt{3}a},\,0)}$, $\displaystyle{\tilde{K}=(-\frac{2\pi}{3a},\,\frac{2\pi}{\sqrt{3}a},\,0)}$ and $a\approx 2.46$\,\AA,
and including sublattices $A$ and $B$ as well as bilayer structure, the four-component wave functions for each valley can be formally written as\,\cite{BilM1}

\begin{equation}
\Psi_K=\left[\begin{array}{c}
\phi_K^A \cr \phi_K^{\tilde{B}} \cr \phi_{K}^{\tilde{A}} \cr \phi_{K}^B
\end{array}\right]\ ,\ \ \ \ \ \ \ \
\Psi_{\tilde{K}}=\left[\begin{array}{c}
\phi_{\tilde{K}}^{\tilde{B}} \cr \phi_{\tilde{K}}^A \cr \phi_{\tilde{K}}^B \cr \phi_{\tilde{K}}^{\tilde{A}}
\end{array}\right]\ ,
\label{e1}
\end{equation}
where $A$ and $B$ label the bonds in the bottom layer and $\tilde{A}$ and $\tilde{B}$ label the bonds in the top layer. For each valley, the $4\times 4$ graphite tight-binding Hamiltonian matrix within the $xy$-plane for Bernal-stacking\,\cite{bern}
bilayer takes the form

\begin{equation}
\hat{\cal H}^{\rm TB}_{\xi}=v_F\left[ {\matrix{ V+\xi u/2 & {\xi v_3(\hat{p}_x+i\hat{p}_y)} & 0 & {\xi v(\hat{p}_x-i\hat{p}_y)} \cr
{\xi v_3(\hat{p}_x-i\hat{p}_y)} & V-\xi u/2 & {\xi v(\hat{p}_x+i\hat{p}_y)} & 0 \cr
0 & {\xi v(\hat{p}_x-i\hat{p}_y)} & V-\xi u/2 & {\gamma_1} \cr
{\xi v(\hat{p}_x+i\hat{p}_y)} & 0 & {\gamma_1} & V+\xi u/2 \cr } } \right]\ ,
\label{e2}
\end{equation}
where $\xi=\pm$ represents the $K$ ($+$) or $\tilde{K}$ ($-$) valley, $v=\displaystyle{\frac{\sqrt{3}}{2\hbar}}\,a\gamma_0\equiv v_F\approx 3\times 10^6$\,cm/s is the intralayer (monolayer) Fermi velocity, $\gamma_1=2m^\ast v^2\ll\gamma_0$ characterizes the effective mass of electrons in the parabolic band, $v_3=\displaystyle{\frac{\sqrt{3}}{2\hbar}}\,a\gamma_1\ll v$ measures the strength of the interlayer coupling, $\displaystyle{\pm\frac{u}{2}}$ represents the bias-induced on-site energies of bilayer, $u=e{\cal E}_0D$ with electric field ${\cal E}_0$ and bilayer separation $D$, and $u=0$ corresponds to a symmetrical bilayer. In addition, we have introduced canonical momentum operators $\hat{p}_x\equiv \displaystyle{-i\hbar\frac{\partial}{\partial x}}+eB_0y$ and
$\hat{p}_y\equiv -i\hbar\displaystyle{\frac{\partial}{\partial y}}$, where the Landau gauge $\mbox{\boldmath$A$}=(-B_0y,\,0,\,0)$ is chosen for a uniform magnetic field ${\bf B}_0$ along the vertical $z$ direction. The potential of a two-dimensional
(2D) scatter array in Eq.\,(\ref{e2}) is assumed as

\begin{equation}
V\equiv V(x,\,y)=V_0\left[\cos\left(\frac{\pi x}{d_x}\right)\,\cos\left(\frac{\pi y}{d_y}\right)\right]^{2N}\ ,
\label{e3}
\end{equation}
where $N$ is an integer, $V_0$ stands for the scattering-potential strength, $d_x$ and $d_y$ are the two array periods in the $x$ and $y$ directions, respectively.
\medskip

Even in the absence of the scatter potential (i.e., $V_0=0$), the eigen-energies and eigen-states correspond to the Hamiltonian in Eq.\,(\ref{e2}) can only be calculated numerically. For low-energy states of electrons
(with kinetic energy less than $\gamma_1/4$), however, the $4\times 4$ Hamiltonian in Eq.\,(\ref{e2}) can be projected onto a $2\times 2$ one. For such a situation, the wave functions in Eq.\,(\ref{e1}) for each valley also reduce
to a two-component form

\begin{equation}
\Psi^K=\left[\begin{array}{c}
\phi_K^A \cr \phi_K^{\tilde{B}} \cr
\end{array}\right]\ ,\ \ \ \ \ \ \ \
\Psi^{\tilde{K}}=\left[\begin{array}{c}
\phi_{\tilde{K}}^{\tilde{B}} \cr \phi_{\tilde{K}}^A \cr
\end{array}\right]\ ,
\label{e4}
\end{equation}
and the projected $2\times 2$ effective Hamiltonian matrix becomes

\[
\hat{\cal H}^{\rm eff}_{\xi}=-\frac{1}{2m^\ast}\left[\begin{array}{cc}
0 & (\hat{p}_x-i\hat{p}_y)^2 \cr
(\hat{p}_x+i\hat{p}_y)^2 & 0 \cr
\end{array}\right]
+\xi v_3\left[\begin{array}{cc}
0 & \hat{p}_x+i\hat{p}_y \cr
\hat{p}_x-i\hat{p}_y & 0 \cr
\end{array}\right]
\]
\begin{equation}
+\frac{\xi u}{2}\left[\begin{array}{cc}
1 & 0 \cr
0 & 1 \cr
\end{array}\right]-\frac{\xi u v^2}{\gamma_1^2}\left[\begin{array}{cc}
(\hat{p}_x-i\hat{p}_y)(\hat{p}_x+i\hat{p}_y) & 0 \cr
0 & -(\hat{p}_x+i\hat{p}_y)(\hat{p}_x-i\hat{p}_y) \cr
\end{array}\right]+V(x,y)\,\hat{I}_0\ ,
\label{e5}
\end{equation}
where $\hat{I}_0$ in the last term is the $2\times 2$ identity matrix, the first, second and the rest two terms represents the intralayer, interlayer and bias effects, respectively.
\medskip

By taking $V=0$ in Eq.\,(\ref{e5}) as a start, in the strong-field limit, i.e., $m^\ast v_3^2\ll\hbar\omega_c<m^\ast v^2$ with a cyclotron frequency $\omega_c=eB_0/m^\ast$, we can formally set $v_3\rightarrow 0$ in Eq.\,(\ref{e5}).
Based on this simplification, we obtain the analytical form of the eigen-energy levels for each valley ($\xi=\pm$)

\begin{equation}
E_{\pm,n}^{\xi}=\left\{\begin{array}{cc}
\pm\hbar\omega_c\sqrt{n(n-1)}-\xi\delta/2\ , & \ \ \ \ \mbox{for $n\geq 2$} \cr
\xi u/2-\xi\delta\ , & \ \ \ \ \mbox{for $n=1$} \cr
\xi u/2\ , & \ \ \ \ \mbox{for $n=0$} \cr
\end{array}\right.\ ,
\label{e6}
\end{equation}
where $\delta=u\,\hbar\omega_c/\gamma_1$, $E_{+,n}^\xi$ and $E_{-,n}^\xi$ correspond to electron and hole energy levels at each valley, respectively, each energy level is spin degenerate, and the lowest two energy levels
are four-fold degenerate with respect to both spins and electron-hole pseudospins. If $u=0$, we get $E_{\pm,0}^\pm=E_{\pm,1}^\pm=0$ from Eq.\,(\ref{e6}),
which becomes eight-fold degenerate now. The corresponding eigen-states to these electron (hole) energy levels ($n\geq 2$) are calculated as

\begin{equation}
\Psi^{K_\xi}_{\pm,n,k_x}(x,\,y)=C^{\pm}_n(\xi)\frac{\displaystyle{e^{i(k_x+K_\xi)x}}}{\sqrt{L_x}}\left[\begin{array}{c}
\phi_{n,k_x+K_\xi}(y) \cr
D^{\pm}_n(\xi)\,\phi_{n-2,k_x+K_\xi}(y) \cr
\end{array}\right]\ ,
\label{e7}
\end{equation}
where $L_x\ (\rightarrow\infty)$ is the sample length in the $x$ direction, $K_\xi=K$ ($K_\xi=\tilde{K}$) for $\xi=+$ ($\xi=-$), $\phi_{n,k_x}(y)\equiv\phi_n(y-y_0)$ is the harmonic-oscillator wave functions with a guiding center $y_0=k_x\ell_B^2$, $\ell_B=\sqrt{\hbar/eB_0}$ the magnetic length, and two coefficients

\begin{equation}
D^{\pm}_n(\xi)=\frac{E_{\pm,n}^{\xi}-\xi u/2+\xi n\delta}{\hbar\omega_c\sqrt{n(n-1)}}\ ,\ \ \ \ \ \ \ \ C^{\pm}_n(\xi)=\frac{1}{\sqrt{1+\left|D^{\pm}_n(\xi)\right|^2}}\ .
\label{e8}
\end{equation}
Assuming $u=0$, we have $D^{\pm}_n(\xi)=\pm 1$ and $C^{\pm}_n(\xi)=1/\sqrt{2}$ for $n\geq 2$, which becomes independent of $\xi$ and $n$. On the other hand, for $n=0$ and $n=1$ we obtain

\[
\Psi^{K_\xi}_{\pm,0,k_x}(x,y)=\frac{{e^{i(k_x+K_\xi)x}}}{\sqrt{L_x}}\left[\begin{array}{c}
\phi_{0,k_x+K_\xi}(y) \cr
0 \cr
\end{array}\right]\ ,
\]
\begin{equation}
\Psi^{K_\xi}_{\pm,1,k_x}(x,y)=\frac{\displaystyle{e^{i(k_x+K_\xi)x}}}{\sqrt{L_x}}\left[\begin{array}{c}
\phi_{1,k_x+K_\xi}(y) \cr
0 \cr
\end{array}\right]\ .
\label{e9}
\end{equation}
\medskip

After the scatter array has been included in the strong-field limit, the wave function of the system can be expanded as

\begin{equation}
\Phi^{\xi}_{\ell;\,\alpha,n,{\bf k}_{\|}} \left( {x,\,y} \right)=\frac{1}{\sqrt{{\cal N}_y}}\sum\limits_{s=-\infty}^{\infty}\left\{{\rm e}^{ik_y\ell_B^2(sp+\ell)K_1}\,\Psi_{\alpha,n,k_x-(sp+\ell)K_1}^{K_\xi} \left( {x,y} \right)\right\}\ ,
\label{e10}
\end{equation}
where $\xi=\pm$, $\alpha=\pm$ corresponds to electron and hole states, $\mbox{\boldmath$k$}_{\|}=(k_x,k_y)$, $|k_x|\leq\pi/d_x=K_1/2$ and $|k_y|\leq\pi/qd_y$ for the first magnetic Brillouin zone,
${\cal N}_y=L_y/(qd_y)$ is the number of unit cells spanned by $b_1=(d_x,\,0)$ and $b_2=(0,\,qd_y)$ in the $y$ direction, $L_y$ ($\to\infty$) is the sample length in the $y$ direction,
$K_1=2\pi/d_x$ is the reciprocal lattice vector in the $x$ direction, and $\ell=1,\,2,\,\cdots,\,p$ is a new quantum number for labeling split $p$ subbands from a $k_x$-degenerated LL in the absence of scatters.
Importantly, the above constructed wave function satisfies the usual Bloch condition, i.e.,

\begin{equation}
\Phi^{\xi}_{\ell;\,\alpha,n,{\bf k}_{\|}} \left( {x+d_x,\,y+qd_y} \right)={\rm e}^{ik_xd_x}\,{\rm e}^{ik_yqd_y}\,\Phi^{\xi}_{\ell;\,\alpha,n,{\bf k}_{\|}} \left( {x,y}\right)\ .
\label{e11}
\end{equation}
Substituting the expression for wave function at each valley in Eq.\,(\ref{e7}) into Eq.\,(\ref{e9}), we find

\[
\Phi^{\xi}_{\ell;\,\alpha,n,{\bf k}_{\|}} \left( {x,y} \right)=\frac{1}{\sqrt{{\cal N}_yL_x}}\sum\limits_{s=-\infty}^{\infty}\left\{{\rm e}^{ik_y\ell_B^2(sp+\ell)K_1}\,C^{\alpha}_n(\xi)\right.
\]
\begin{equation}
\left.\times e^{i[k_x+K_\xi-(sp+\ell)K_1]x}\left[\begin{array}{c}
\phi_{n,k_x+K_\xi-(sp+\ell)K_1}(y) \cr
D^{\alpha}_n(\xi)\,\phi_{n-2,k_x+K_\xi-(sp+\ell)K_1}(y) \cr
\end{array}\right]\right\}\ ,
\label{e12}
\end{equation}
where $D^{\alpha}_0(\xi)=D^{\alpha}_1(\xi)=0$ and $C^{\alpha}_0(\xi)=C^{\alpha}_1(\xi)=1$.
\medskip

Now, by taking into account of the $V(x,y)\hat{I}_0$ term in Eq.\,(\ref{e5}), a tedious but straightforward calculation leads to an explicit expression for the matrix elements of the potential $V(x,y)$, yielding

\[
V^{\ell',n',\alpha'}_{\ell,n,\alpha}(\mbox{\boldmath$k$}_{\|},\xi)\equiv\sum\limits_{\xi',{\bf k}'_{\|}}
\int\int\,dxdy\,\left[\Phi^{\xi'}_{\ell';\,\alpha',n',{\bf k}'_{\|}} \left( {x,y} \right)\right]^\dag\,V(x,\,y)\,\Phi^{\xi}_{\ell;\,\alpha,n,{\bf k}_{\|}} \left( {x,\,y} \right)
\]
\[
=\frac{V_0}{4^{2N}}\sum\limits_{\xi'}\,C^{\alpha'}_{n'}(\xi')C^{\alpha}_n(\xi)\left\{{\rm e}^{ik_y\ell_B^2K_1(\ell-\ell')}\,\sum\limits_{i=0}^{N-1}\sum\limits_{j=0}^{N-1}\,\left[{\cal F}^{(B)}_{ij}(\xi',\xi)+D^{\alpha'}_{n'}(\xi')D^{\alpha}_n(\xi)\,{\cal F}^{(A)}_{ij}(\xi',\xi)\right]\right.
\]
\begin{equation}
\left.+\delta_{\ell,\ell'}\delta_{n,n'}\delta_{\xi,\xi'}\,\left[1+D^{\alpha'}_{n}(\xi)D^{\alpha}_n(\xi)\right]\left[\frac{(2N)!}{(N!)^2}\right]^2\right\}\ ,
\label{e13}
\end{equation}
where $\alpha,\,\alpha'=\pm$ correspond to electron and hole levels, respectively. From Eq.\,(\ref{e13}) we find two valleys for bilayer graphene can be coupled to each other, which is different from the monolayer graphene\,\cite{ourRK}.
Here, the terms with $\xi'=\xi$ come from the intravalley contribution, whereas the terms with $\xi'\neq\xi$ stand for the intervalley coupling which presents an {\em interference} effect. Moreover, we have defined in Eq.\,(\ref{e13}) two
intervalley ($\xi\neq\xi'$) coupling factors

\[
{\cal F}_{ij}^{(B,A)}(\xi',\xi)=\left( {\matrix{ {2N} \cr {i} \cr } } \right)\left( {\matrix{ {2N} \cr {N} \cr } } \right)A^{(B,A)}_1(0,\,N-i\left|\xi',\xi\right.)
\]
\begin{equation}
+\left( {\matrix{ {2N} \cr {j} \cr } } \right)\left( {\matrix{ {2N} \cr {N} \cr } } \right)A^{(B,A)}_2(N-j,\,0)
+2\left( {\matrix{ {2N} \cr {i} \cr } } \right)\left( {\matrix{ {2N} \cr {j} \cr } } \right)A^{(B,A)}_3(N-j,\,N-i\left|\xi',\xi\right.)\ ,
\label{e14}
\end{equation}
where the binomial expansion coefficient for $m\geq n$ is

\begin{equation}
\left( {\matrix{ {m} \cr {n} \cr } } \right)\equiv\frac{m!}{n!\,(m-n)!}\ .
\label{e15}
\end{equation}
Finally, we have introduced in Eq.\,(\ref{e14}) the following three self-defined functions

\begin{equation}
A_1^{(B,A)}(r,\,s\left|\xi',\xi\right.)=D_{n',n}^{rs{(B,A)}}\,T_{\ell}^s(\xi',\xi)\,\delta_{\ell,\ell'}\ ,
\label{e16}
\end{equation}

\begin{equation}
A^{(B,A)}_2(r,\,s)=D_{n',n}^{rs{(B,A)}}\left\{\delta_{\ell-\ell',r}\left[{\rm sgn}(n'-n)\right]^{\beta}+\delta_{\ell'-\ell,r}\left[{\rm sgn}(n-n')\right]^{\beta}\right\}\ ,
\label{e17}
\end{equation}

\[
A^{(B,A)}_3(r,\,s\left|\xi',\xi\right.)=D_{n',n}^{rs{(B,A)}}\left\{\delta_{\ell-\ell',r}\left[{\rm sgn}(n'-n)\right]^{\beta}\cos[\Theta_{rs}^{\ell'}(n',\,n\left|\xi',\xi\right.)]\right.
\]
\begin{equation}
\left.+\delta_{\ell'-\ell,r}\left[{\rm sgn}(n-n')\right]^{\beta}\cos[\Theta_{rs}^{\ell}(n,\,n'\left|\xi',\xi\right.)]\right\}\ ,
\label{e18}
\end{equation}
where $\beta=|n-n'|$,

\begin{equation}
D_{n',n}^{rs(B)}=\sqrt{\frac{n_1!}{n_2!}}\,{\rm e}^{-W_{rs}/(2\phi)}\left(\frac{W_{rs}}{\phi}\right)^{\beta/2}L_{n_1}^{(\beta)}\left(\frac{W_{rs}}{\phi}\right)\ ,
\label{e19}
\end{equation}
$\phi\equiv\Phi/\Phi_0=p/q$ with $p$ and $q$ being the integers prime to each other, $\Phi=B_0d_xd_y$ is the magnetic flux per unit cell, $\Phi_0=h/e$ is the flux quanta,
$n_1={\rm min}(n,\,n')$, $n_2={\rm max}(n,\,n')$, $L_n^{(m)}(x)$ is the associated Laguerre polynomial, $W_{rs}=\pi(r^2K_1^2+s^2K_2^2)/K_1K_2$, $K_2=2\pi/d_y$, $D_{n',n}^{rs(A)}=D_{n'-2,n-2}^{rs(B)}$,

\[
T_\ell^s(\xi',\xi)
\]
\begin{equation}
=\left\{
\begin{array}{ll}
\pm 2\cos\left[\displaystyle{\frac{s(\tilde{k}_x(\xi',\xi)d_x-\ell 2\pi)}{\phi}}\right]\ , & \mbox{$(+)$ for $\beta=4N$ and $(-)$ for $\beta=4N+2$}\\
\\
\pm 2\sin\left[\displaystyle{\frac{s(\tilde{k}_x(\xi',\xi)d_x-\ell 2\pi)}{\phi}}\right]\ , & \mbox{$(+)$ for $\beta=4N+1$ and $(-)$ for $\beta=4N+3$}
\end{array}\ ,\right.
\label{e20}
\end{equation}

\begin{equation}
\Theta^\ell_{rs}(n',\,n\left|\xi',\xi\right.)=\frac{s[\tilde{k}_x(\xi',\xi)d_x-2\pi\left(\ell+r/2\right)]}{\phi}-{\rm sgn}(n'-n)\,\beta\tan^{-1}\left(\frac{sd_x}{rd_y}\right)\ ,
\label{e21}
\end{equation}
and $\tilde{k}_x(\xi',\xi)=k_x+(K_\xi-K_{\xi'})$ characterizing the intervalley coupling for $\xi'\neq\xi$ and the interference effect as well. Here, the range of $k_x$ extends to all magnetic Brillouin zones in this direction for Umklapp scatterings.
\medskip

\begin{figure}
\centering
\includegraphics[width=0.99\textwidth]{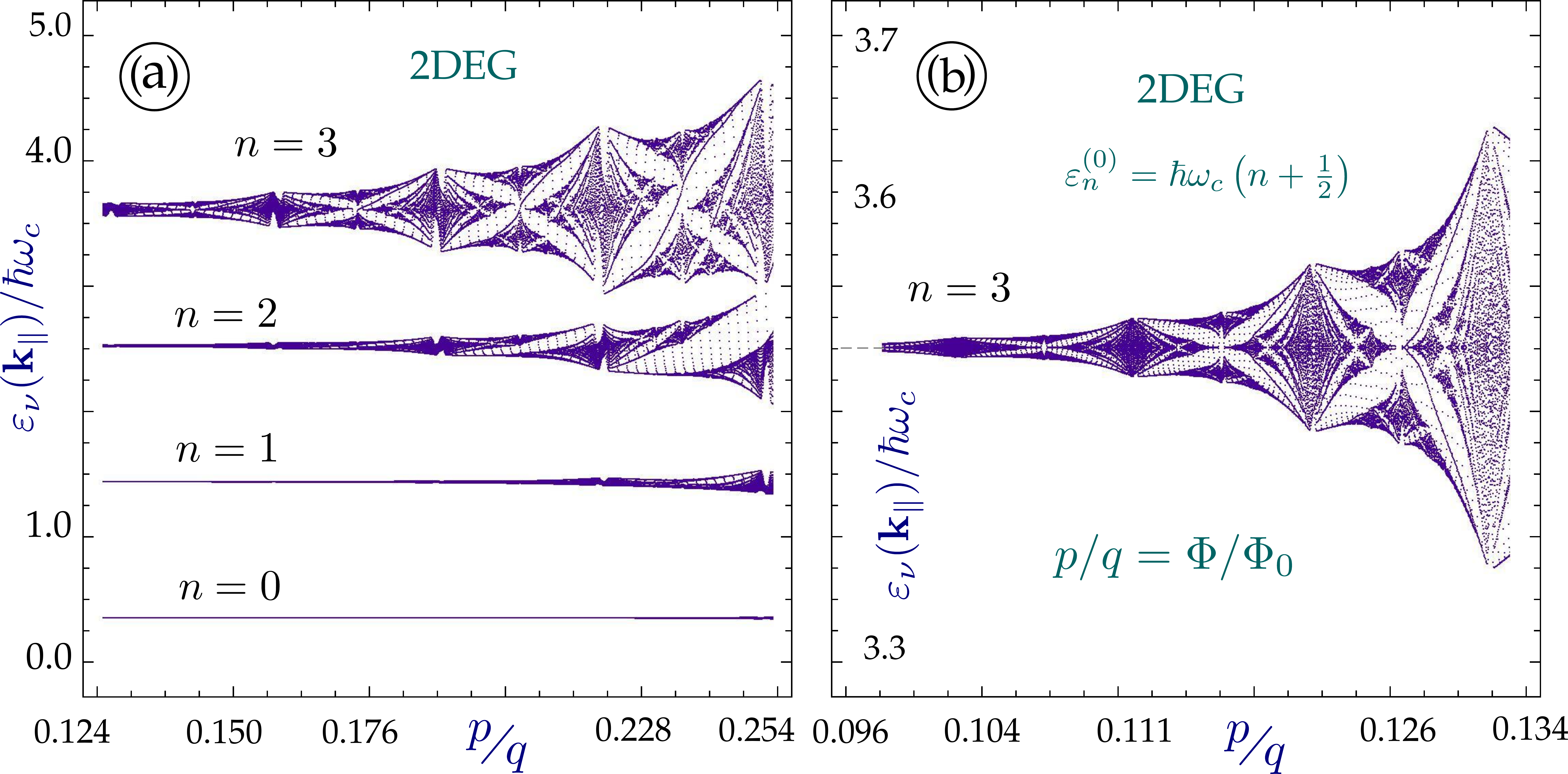}
\caption{(Color online) Distributions of magnetically-quantized energy levels $\varepsilon_\nu(\mbox{\boldmath$k$}_{\|})$ of a 2DEG as a function of magnetic flux $\Phi/\Phi_0=p/q$
under a 2D scattering-lattice potential given by Eq.\,(\ref{e3}) with parameters $V_0/\hbar\omega_c= 1$, $N=3$, $d_x=d_y$, $\omega_c=eB_0/m^\ast$, and $m^\ast$ as the effective mass of electrons.
Here, we have chosen $k_x=k_y = 0.3\,K_1$. Panel $(a)$ displays the distributions of the lowest four bands,
and panel $(b)$ shows close-up view of the self-similar pattern of the $n=3$ band at lower $B_0$.}
\label{FIG:1}
\end{figure}

The energy dispersion $\varepsilon_{\nu}(\mbox{\boldmath$k$}_{\|},\xi)$ of the $\nu$th magnetic band
around each valley for this modulated system is a solution of the eigenvector problem $\tensor{\mbox{\boldmath${\cal M}$}}(\mbox{\boldmath$k$}_{\|},\xi)\cdot\mbox{\boldmath${\cal A}$}(\mbox{\boldmath$k$}_{\|},\xi)=0$ with elements of the coefficient
matrix $\tensor{\mbox{\boldmath${\cal M}$}}(\mbox{\boldmath$k$}_{\|},\xi)$ given by

\begin{equation}
\{\tensor{\mbox{\boldmath${\cal M}$}}(\mbox{\boldmath$k$}_{\|},\xi)\}_{j,\,j'}=\left[E^{\xi}_{\alpha,\,n}-\varepsilon(\mbox{\boldmath$k$}_{\|},\xi)\right]
\delta_{n,n'}\delta_{\ell,\ell'}\delta^{(n)}_{\alpha,\alpha'}+V^{\ell',n',\alpha'}_{\ell,n,\alpha}(\mbox{\boldmath$k$}_{\|},\xi)\ ,
\label{e22}
\end{equation}
where $\delta^{(n)}_{\alpha,\alpha'}=1$ for $n=0,\,1$ (i.e., degenerate electron-hole levels) and $\delta^{(n)}_{\alpha,\alpha'} =
\delta_{\alpha,\alpha'}$ for $n\geq 2$, $j=\{n,\,\ell,\,\alpha\}$ is a composite index, and $\{\mbox{\boldmath${\cal A}$}(\mbox{\boldmath$k$}_{\|},\xi)\}_j \equiv
{\cal A}^{\alpha}_{n,\ell}(\mbox{\boldmath$k$}_{\|},\xi)$ is an orthonormal eigenvector. Furthermore, the eigenvalues $\varepsilon(\mbox{\boldmath$k$}_{\|},\xi)$ of the system are
determined by roots of the characteristic equation ${\rm Det}\,\tensor{\mbox{\boldmath${\cal M}$}}(\mbox{\boldmath$k$}_{\|},\xi)=0$.

\section{Numerical Results and Discussions}
\label{s3}

\subsection{Two-Dimensional Electron Gas and Monolayer Graphene}

As a starting point, we first briefly discuss the effect of a two-dimensional (2D) periodically-modulated scattering-lattice potential in Eq.\,(\ref{e3}) on a 2D electron gas (EG) under a perpendicular quantizing magnetic field $B_0$.
In the absence of this scattering-lattice potential, 2DEG will be quantized into a series of discrete
LLs: $\varepsilon^{(0)}_n=(n+1/2)\,\hbar\omega_c$ with $n=0,\,1,\,2,\,\cdots$, $\omega_c=eB_0/m^\ast$ as the cyclotron frequency, and $m^\ast$ as the effective mass of electrons.
These uncoupled LLs are highly degenerate with respect to their guiding centers $y_0=k_x\ell^2_B$ (or with different cyclotron orbits),
where $\ell_B=\sqrt{\hbar/eB_0}$ is the magnetic length. In the presence of the scattering-lattice potential, however, these degenerate LLs are strongly coupled to each other and expand into a set of split Landau bands,
as shown in Fig.\,\ref{FIG:1}$(a)$. Furthermore, a close-up view in Fig.\,\ref{FIG:1}$(b)$ reveals that a self-similar pattern occurs within the fourth ($n=3$) Landau band at low $B_0$, just as predicted early by Hofstadter
in his seminal work\,\cite{hofs-main}.
\medskip

\begin{figure}
\centering
\includegraphics[width=0.99\textwidth]{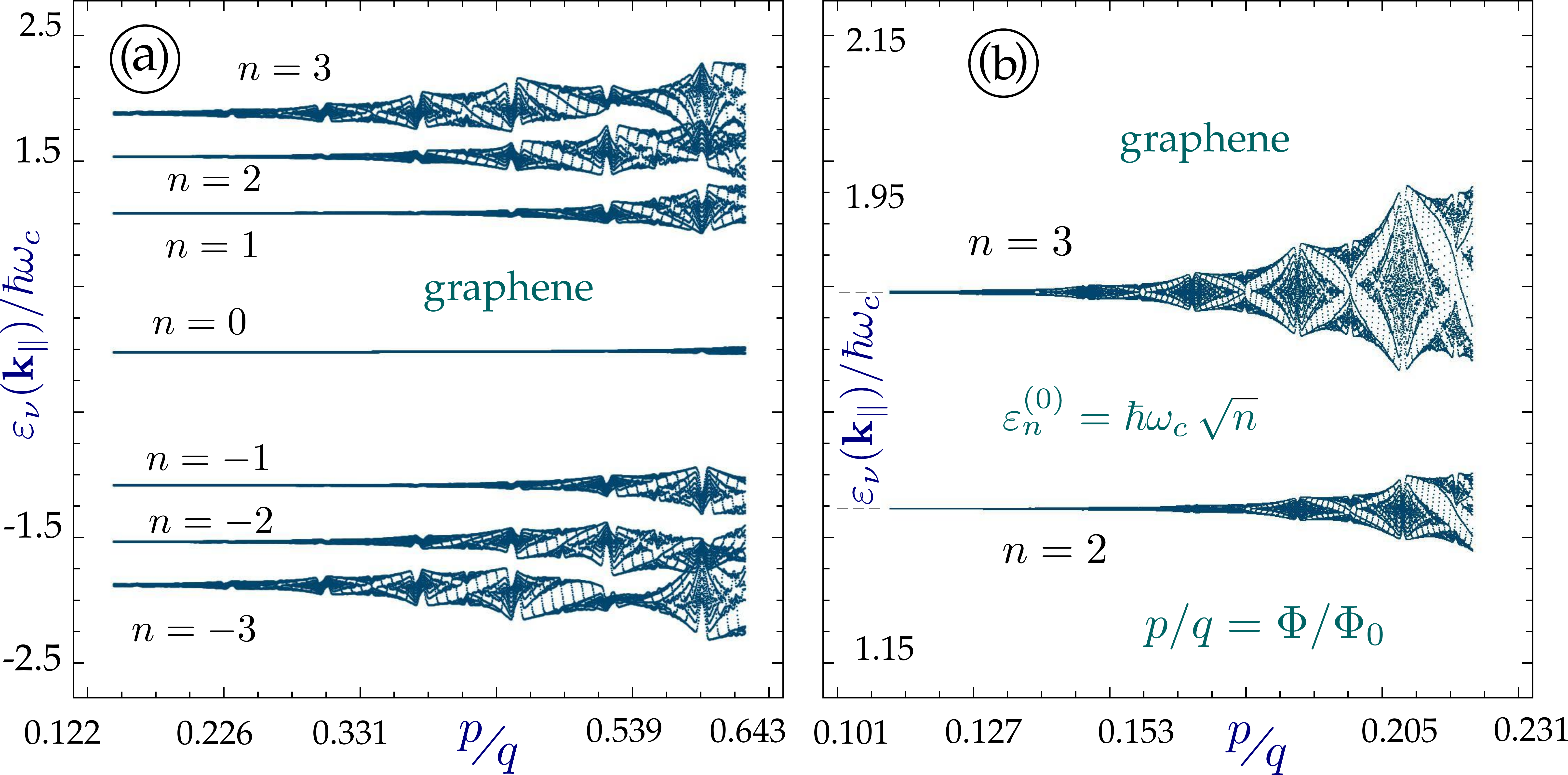}
\caption{(Color online) Distributions of quantized energy levels $\varepsilon_\nu(\mbox{\boldmath$k$}_{\|})$ of a monolayer graphene as functions of $\Phi/\Phi_0 = p/q$
under the same 2D scattering-lattice potential in Eq.\,(\ref{e3}) with parameters $V_0/\hbar\omega_c= 1$, $N=3$, $d_x=d_y$, $\omega_c=\sqrt{2}v_F/\ell_B$, $\ell_B=\sqrt{\hbar/eB_0}$ as the magnetic length, and
$v_F$ the Fermi velocity.
Here, we take $k_x=k_y = 0.3\,K_1$.
Panel $(a)$ presents the distributions of the lowest four bands for electrons and holes,
while panel $(b)$ highlights close-up view of the self-similar structures of the $n=2$ and $n=3$ electron Landau bands at lower $B_0$.}
\label{FIG:2}
\end{figure}

\begin{figure}
\centering
\includegraphics[width=0.5\textwidth]{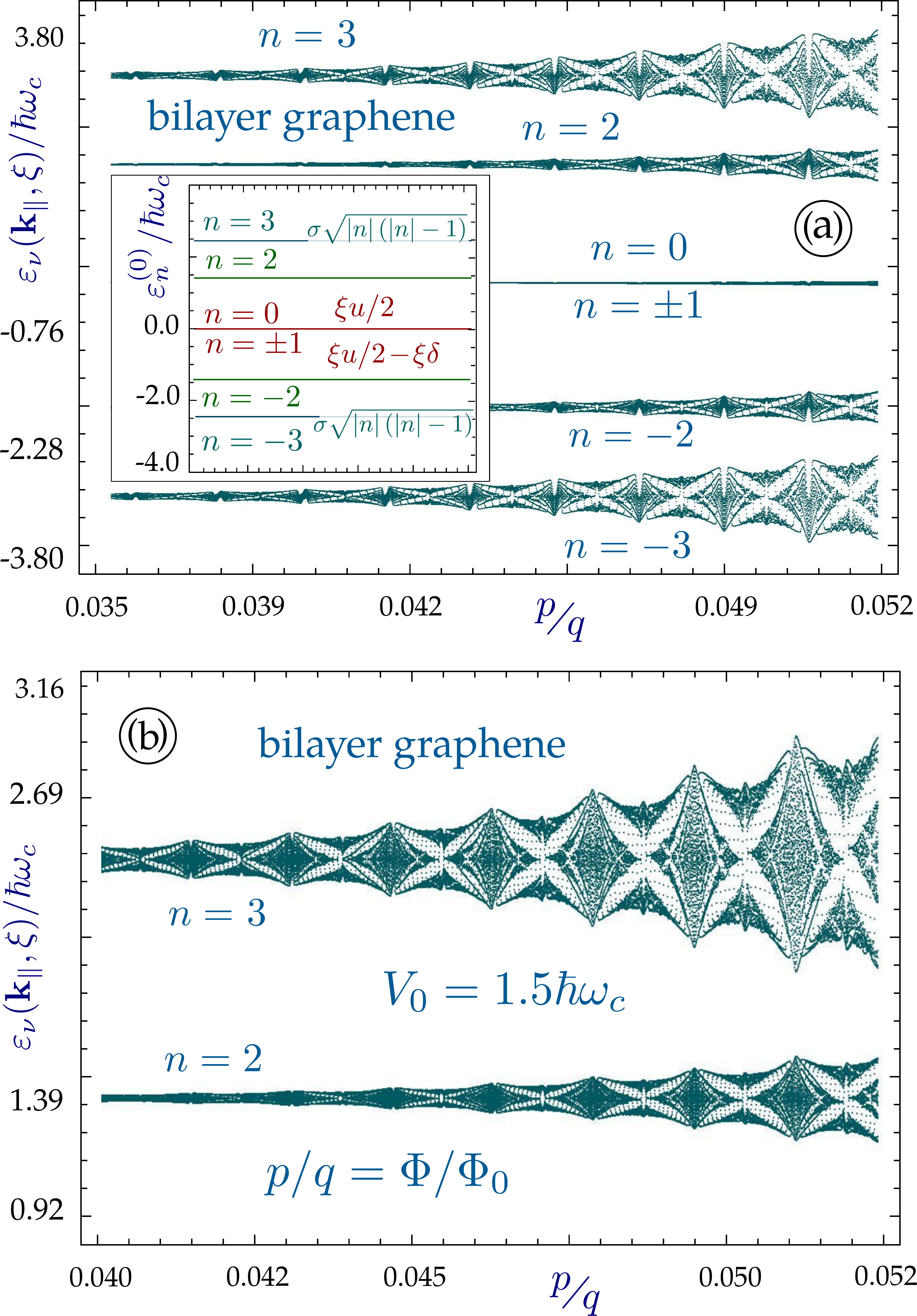}
\caption{(Color online) Distributions of energy levels $\varepsilon_\nu(\mbox{\boldmath$k$}_{\|},\xi)$ of a bilayer graphene as functions of $\Phi/\Phi_0 = p/q$
under the same 2D scattering-lattice potential in Eq.\,(\ref{e3}) with parameters $\xi=+$, $\delta/\hbar\omega_c=0.001$, $u/\hbar\omega_c=0.003$,
$V_0/\hbar\omega_c = 1.5$, $N=3$, $d_x=d_y=2.46\,nm$, $\omega_c = e B_0/m^*$, $m^\ast$ as the effective mass of electrons or holes, and
$\alpha = \pm 1$ is the pseudospin index for electrons ($+$) and holes ($-$), respectively. Here, we set
$k_x=k_y = 0.3\,K_1$. Panel $(a)$ presents the distributions
of the lowest four Landau bands for electrons and holes, and panel $(b)$ displays the close-up view of the self-similar patterns of the $n=2$
and $n=3$ electron Landau bands.}
\label{FIG:3}
\end{figure}

If the 2DEG is replaced by a monolayer graphene, a different set of LLs $\varepsilon_{n,\pm}^{(0)}={\rm sgn}(n)\,\hbar\omega_c\sqrt{|n|}$ with $n=0,\,\pm 1,\,\pm 2,\,\cdots$ appears in the absence of a scattering-lattice potential,
where $\omega_c=\sqrt{2}v_F/\ell_B$, $v_F$ is the Fermi velocity of graphene, and $+$ ($-$) corresponds to electrons (holes), respectively. In this case, we find that the $n=0$ LL sits at the zero-energy Dirac point
instead of $\hbar\omega_c/2$ for 2DEG, and
$\varepsilon^{(0)}_{n,\pm}\propto\sqrt{|n|B_0}$ but not proportional to $(n+1/2)B_0$ for 2DEG.
After the scattering-lattice potential in Eq.\,(\ref{e3}) has been employed, these guiding-center degenerated
energy levels also expand into a Landau band through mutual couplings, as seen in Fig.\,\ref{FIG:2}$(a)$. However,
the mirror symmetry with respect to the band center is lost in Fig.\,\ref{FIG:2}$(b)$ for monolayer graphene, as discussed in details recently by us\,\cite{ourRK}.
Here, one crucial difference between 2DEG and monolayer graphene is the LL separation $(\sqrt{n+1} - \sqrt{n})\,\hbar\omega_c$ for graphene, in contrast with a uniform one, $\hbar\omega_c$, for 2DEG.
Consequently, the graphene energy-level separation will decrease with increasing $n$, and therefore, overlaps of many Hofstadter butterflies will show up for higher $n$ values as in Fig.\,\ref{FIG:2}$(a)$.

\begin{figure}
\centering
\includegraphics[width=0.99\textwidth]{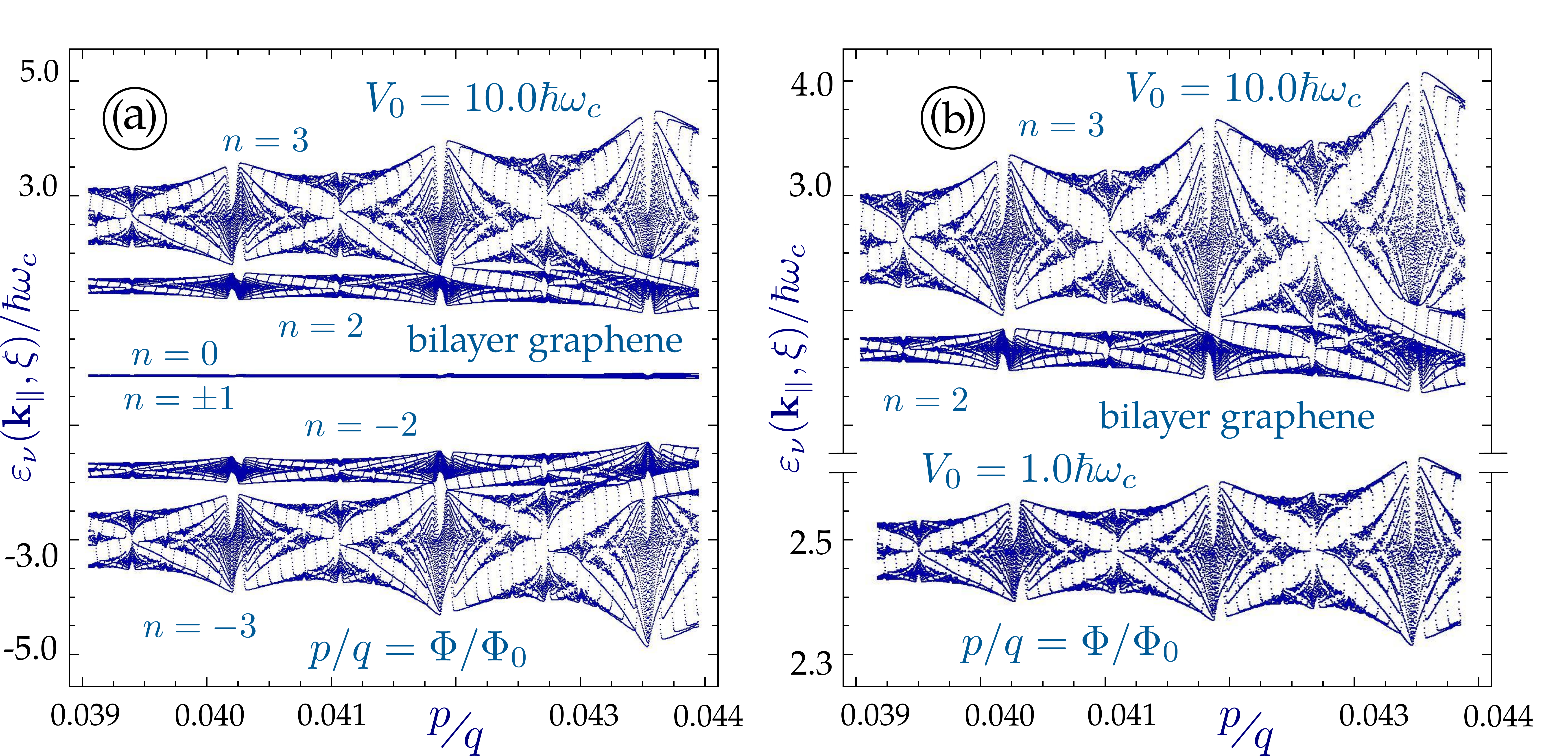}
\caption{(Color online) Distributions of energy levels $\varepsilon_\nu(\mbox{\boldmath$k$}_{\|},\xi)$ of a bilayer graphene as functions of $\Phi/\Phi_0= p/q$ under the same 2D scattering-lattice potential in Eq.\,(\ref{e3}) with a strong modulation $V_0/\hbar\omega_c = 10$.	
The other parameters are the same as those in Fig.\,\ref{FIG:3}.
Panel $(a)$ presents the distributions
of the lowest four Landau bands for electrons and holes, while panel $(b)$ displays a close-up view for a comparison of the self-similar patterns within the $n=2$ electron
Landau band at $V_0/\hbar\omega_c = 10$ and $V_0/\hbar\omega_c = 1$, respectively.}
\label{FIG:4}
\end{figure}

\begin{figure}
\centering
\includegraphics[width=0.9\textwidth]{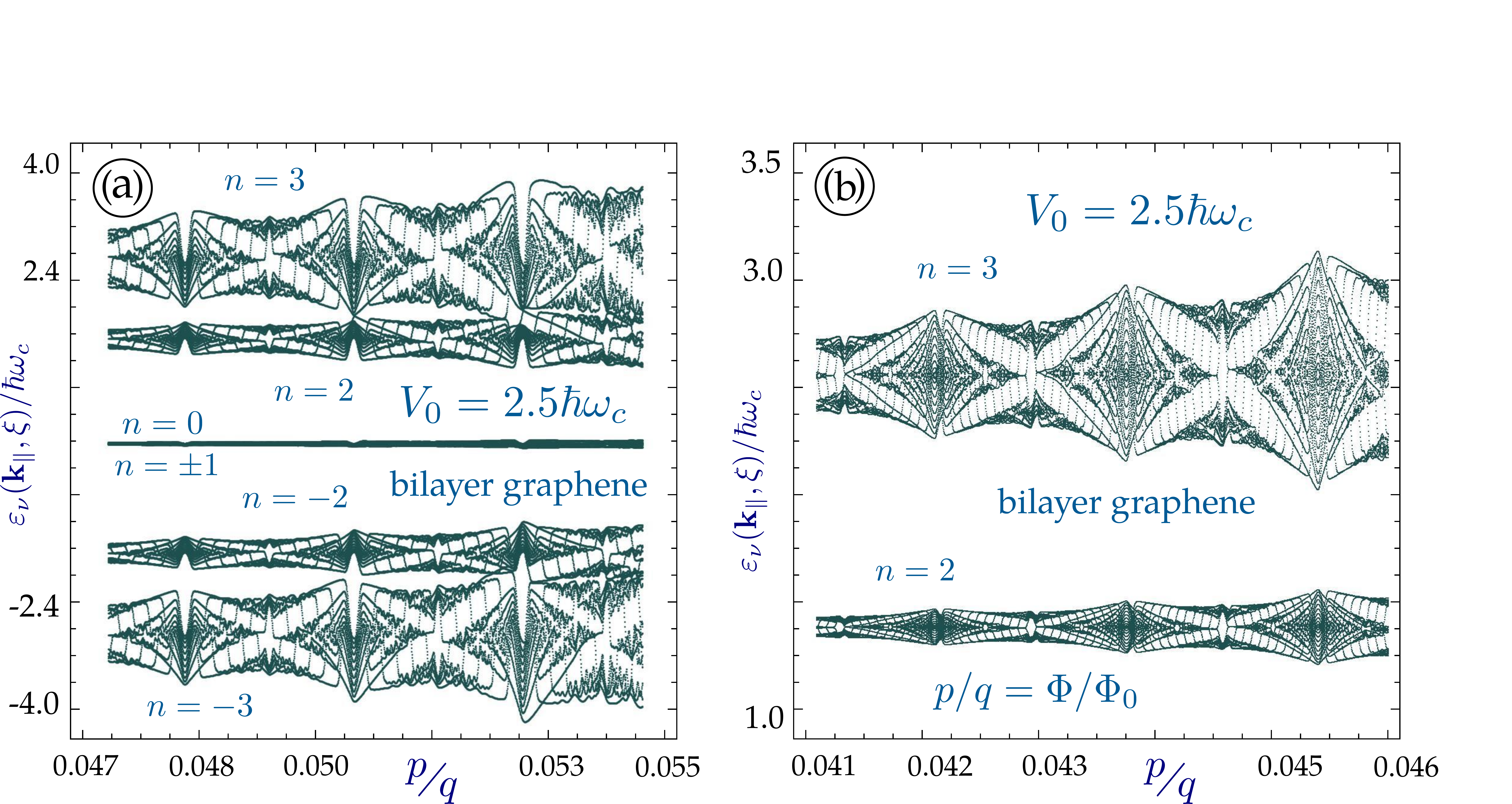}
\caption{(Color online) Distributions of energy levels $\varepsilon_\nu(\mbox{\boldmath$k$}_{\|},\xi)$ of a bilayer graphene as functions of $\Phi/\Phi_0 = p/q$
under the same 2D scattering-lattice potential in Eq.\,(\ref{e3}) with an intermediate modulation $V_0/\hbar\omega_c = 2.5$.
The other parameters are the same as those in Fig.\,\ref{FIG:3}.
Panel $(a)$ presents the distributions
of the lowest four Landau bands for electrons and holes, while panel $(b)$ displays a close-up view of the self-similar structures of the $n=2$ and $n=3$ electron Landau bands at lower $B_0$.}
\label{FIG:5}
\end{figure}

\begin{figure}
\centering
\includegraphics[width=0.5\textwidth]{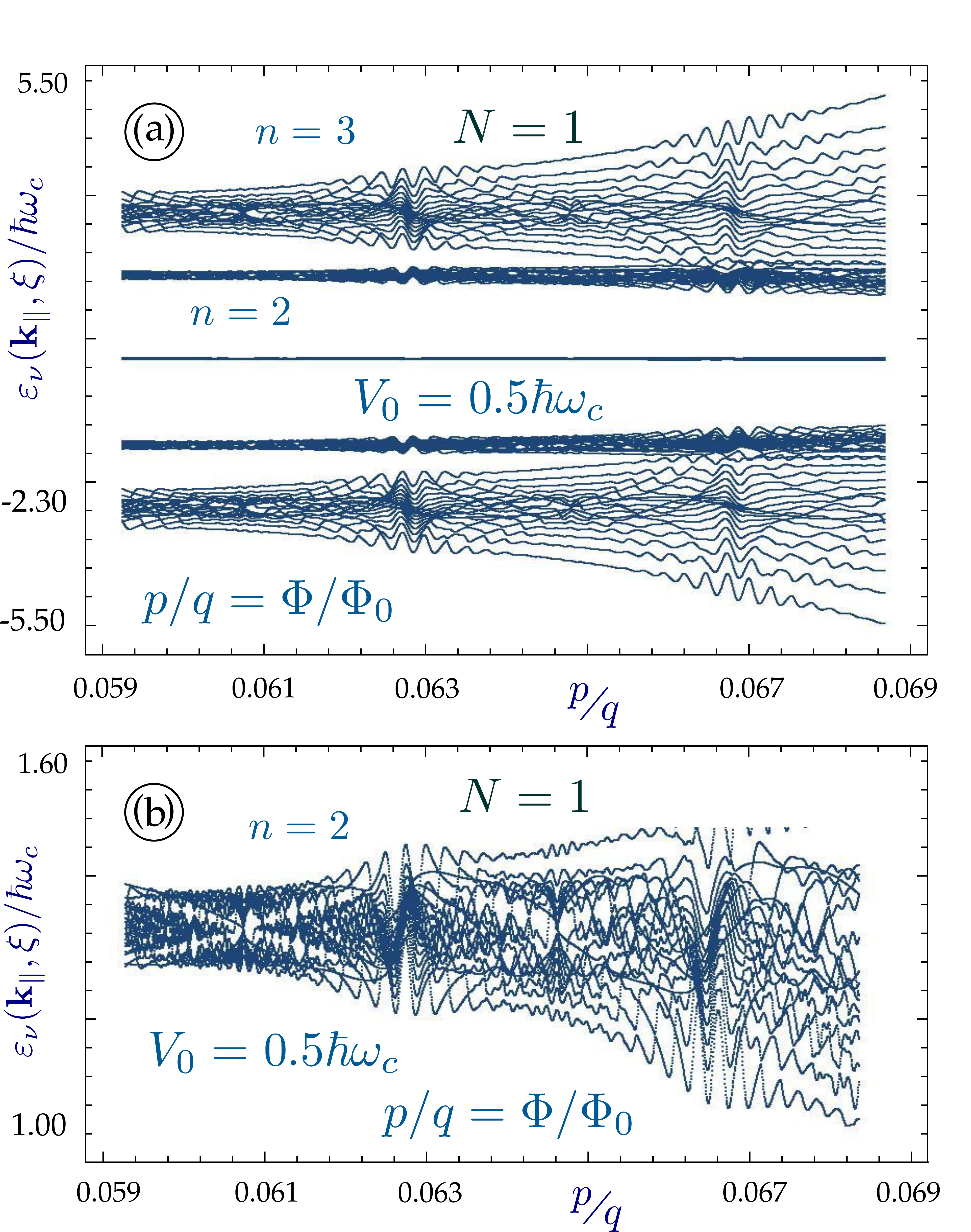}
\caption{(Color online) Distributions of energy levels $\varepsilon_\nu(\mbox{\boldmath$k$}_{\|},\xi)$ of a bilayer graphene as functions of $\Phi/\Phi_0 = p/q$
under the same 2D scattering-lattice potential in Eq.\,(\ref{e3}) with a weak modulation $V_0/\hbar\omega_c = 0.5$ for a weaker bias field ${\cal E}_0$ with $\delta/\hbar\omega_c=0.0025$ and $u/\hbar\omega_c=0.0075$.
The other parameters are the same as those in Fig.\,\ref{FIG:3} except for $N=1$.		
Panel $(a)$ presents the distributions
of the lowest four Landau bands for electrons and holes, and panel $(b)$ displays the close-up view of the significantly-deformed self-similar patterns of the $n=2$
electron Landau band.}
\label{FIG:6}
\end{figure}

\begin{figure}
\centering
\includegraphics[width=0.5\textwidth]{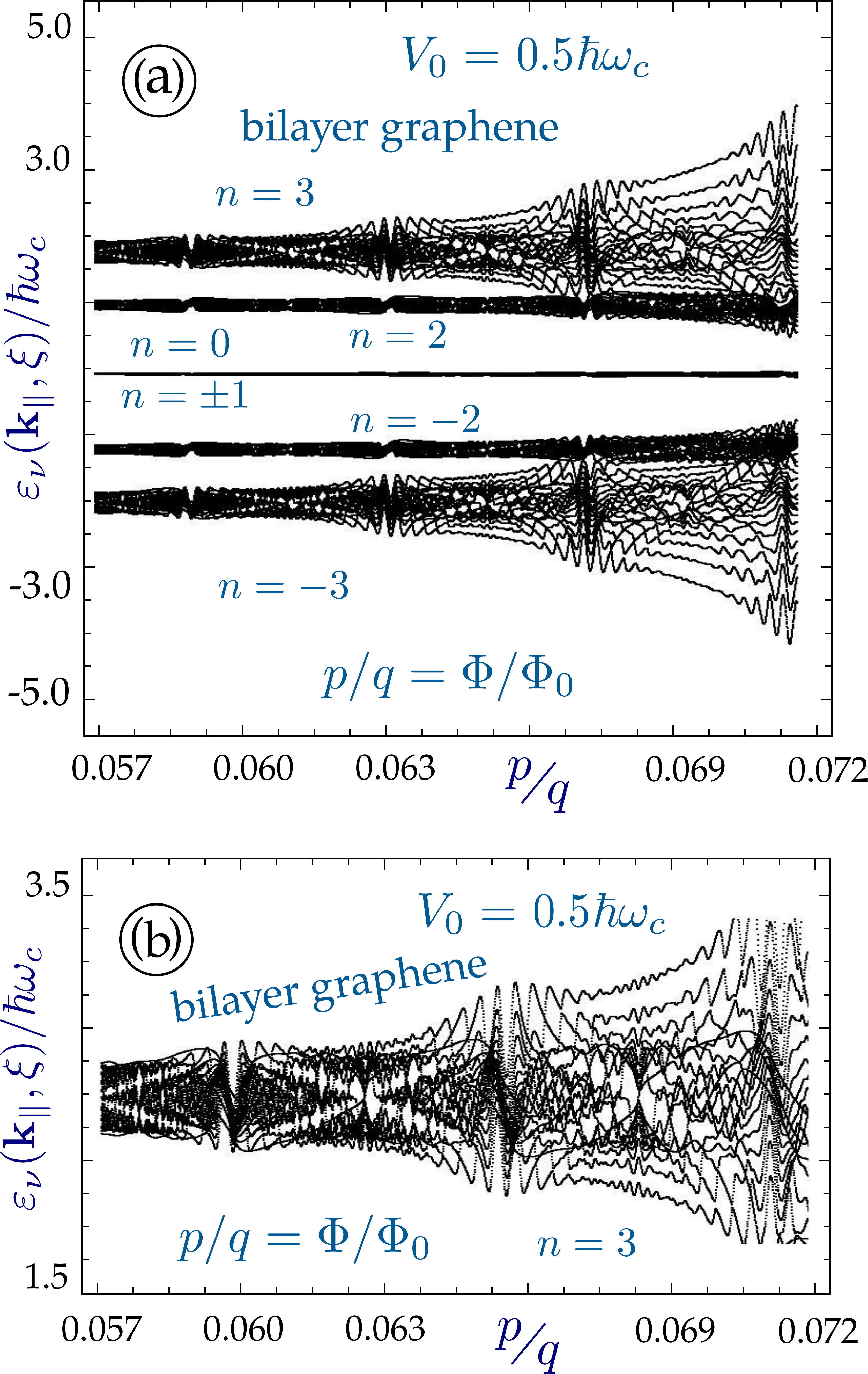}
\caption{(Color online) Distributions of energy levels $\varepsilon_\nu(\mbox{\boldmath$k$}_{\|},\xi)$ of a bilayer graphene as functions of $\Phi/\Phi_0 = p/q$
under the same 2D scattering-lattice potential in Eq.\,(\ref{e3}) with a weak modulation $V_0/\hbar\omega_c = 0.5$ for a very-strong bias field ${\cal E}_0$ with $\delta/\hbar\omega_c=0.1$ and $u/\hbar\omega_c=0.3$.
The other parameters are the same as those in Fig.\,\ref{FIG:3}.		
Panel $(a)$ presents the distributions
of the lowest four Landau bands for electrons and holes, and panel $(b)$ displays the close-up view of the completely-destroyed self-similar patterns of the $n=3$
electron Landau band.}
\label{FIG:7}
\end{figure}

\subsection{Bilayer Graphene}

Now, Let us turn our attention to discussions on development of Landau bands in a bilayer graphene.
For bilayer graphene subjected to a scattering-lattice potential given by Eq.\,(\ref{e3}) and under a perpendicular quantizing magnetic field $B_0$ at the same time,
our numerical solutions for the eigenvalue equation in Eq.\,(\ref{e22}) are presented in Figs.\,\ref{FIG:3} - \ref{FIG:5} with various scattering strengths $V_0$. As a whole, we find that
degenerate LLs with different guiding centers tend to couple to each other and lead to band-center asymmetric Landau bands within which a fractal Hostadter structure is seen for high magnetic fields $B_0$.
Furthermore, the developed Landau bands for two valleys ($\xi=\pm$) are coupled to each other in a bilayer graphene through an Umklapp scattering process across whole magnetic Brillouin zones, which is in contrast with the case for a monolayer graphene
where the Landau bands are found independent of a valley.
\medskip

As indicated in Section\ \ref{s2}, the valley mixing and interference effect contained in the modulation potential
$V^{\ell',n',\alpha'}_{\ell,n,\alpha}(\mbox{\boldmath$k$}_{\|},\xi)$ in Eq.\,(\ref{e13}) are described explicitly by the wave number $\tilde{k}_x(\xi',\xi) = k_x +
(K_\xi-K_{\xi'})$, where  $K_\xi = - K_{\xi'} = 20.94\, d_x^{-1}$, and $d_x = d_y = 10 a = 2.46\,$nm. The integer power $N$, which measures the peak sharpness of the scattering potential in Eq.\,(\ref{e3}),
is selected as $N=3$. In the absence of the 2D scattering-lattice potential, each LL under the magnetic flux ratio $\Phi/\Phi_0=p/q$ has a $p$-fold degeneracy for magnetic subbands.
We have taken $p=13$, $17$ and $11$, respectively, in Figs.\,\ref{FIG:3}$-$\ref{FIG:5}. For all three
graphs, we only show the lowest four Landau bands for both electrons and holes. All the numerical results which display self-repeated Hofstadter butterfly structures are presented
as a function of $\Phi/\Phi_0 = p/q$.
Here, all energy levels, except for $n=0$ and $n=1$, are shifted upwards by a fixed energy offset $4^{-2N}\,\{ (2N)!/N! \}^2 \, V_0 = 0.146\,V_0 $ for $N=3$. Therefore,
we have to made an adjustment to our plots in Figs.\,\ref{FIG:3}$-$\ref{FIG:5}
so that the electron-hole symmetry can be restored with respect to the zero-energy point.
With fixed lattice period $d_x=d_y$, a magnetic-flux ratio $\Phi/\Phi_0 = p/q$ can be uniquely related to a magnetic-field strength $B_0$.
The upper bound of $p/q$ in Figs.\,\ref{FIG:3}$-$\ref{FIG:5} for observing Hofstadter spectra is found within the range of $B_0=5-10\,T$.
\medskip

The unperturbed LL spectrum is shown in Eq.\,(\ref{e6}). In our numerical calculations, we have set $\delta/\hbar\omega_c\approx 0.001$ and
$u/\hbar\omega_c\approx 0.003$ so that the LL structure consists of a few pairs of extremely closely-located levels, corresponding to $\xi,\,\xi'=
\pm$ for two valley indexes. This on-site energy-level separation ($\sim 10^{-3}\,\hbar \omega_c$) depends on $B_0$ or
$p/q$. Additionally, two groups of LaLs associated with $n=0$ and $n=\pm 1$ are nearly degenerate due to their very small separations $\delta$, as found from the inset of Fig.\,\ref{FIG:3}$(a)$.
Furthermore, the spin degeneracy in these LLs is kept since none of them depends
on spin index. All $\hbar\omega_c-$scaled higher levels staring for $n\geq 2$ have the same $ \alpha\, \sqrt{n(n-1)}$ dependence which becomes nearly
equidistant as $n \gg 1$ and in contrast with the monolayer graphene. Here, the pseudospin index $\alpha = \pm 1$ hints a complete electron/hole symmetry for these
$n \geq 2$ LLs. After the scattering-lattice potential given by Eq.\,(\ref{e3}) has been introduced to bilayer graphene,
the previously uncoupled and highly-degenerate LLs expand into many magnetic bands with self-similar structures, as
can be verified directly from Fig.\,\ref{FIG:3}$(b)$.
Since the higher LLs become almost equally separated in bilayer graphene, we expect similar self-repeated structures within a magnetic band for large $n$ values.
\medskip

Because the mixing of LaLs depends on $V_0$, we present comparisons in Figs.\,\ref{FIG:4} and \ref{FIG:5} for strong and intermediate scattering strengths $V_0/\hbar\omega_c$.
When the strong scattering strength is $V_0/\hbar\omega_c=10$, the mixing of $n=2$ and $n=3$ Landau bands is severe, as seen in Fig.\,\ref{FIG:4}$(a)$.
In addition, the band mixing is found to increase with magnetic field $B_0$ in this case. If the scattering strength, $V_0/\hbar\omega_c=1$, is weak, on the other hand, no band mixing appears, as can be verified from Fig.\,\ref{FIG:4}$(b)$.
\medskip

For intermediate scattering strength $V_0/\hbar\omega_c=2.5$ in Fig.\,\ref{FIG:5}$(a)$, we find the band mixing still happen, but it occurs at a higher magnetic field.
For lower values of $B_0$, on the other hand, such band mixing is completely negligible, as found from Fig.\,\ref{FIG:5}$(b)$.
Therefore, in order to observe Hofstadter butterflies and band mixing effects simultaneously, a stronger scattering strength $V_0$ is preferred. More importantly, the large value of $V_0/\hbar\omega_c$
also brings down the required magnetic field for observation to an experimentally accessible level.

\subsection{Effect of Breaking Down of Inversion Symmetry}

For a monolayer graphene, the group of wavevector associated with the $K$ or $K'$ point within the crystal first Brillouin zone is found isomorphic to the point group\,\cite{pointg} $D_{3h}$. For a bilayer graphene with a Bernal stacking, on the other hand, this $D_{3h}$ point group is downgraded to $D_3$ with a lower symmetry. Furthermore, in the presence of a vertical bias field, these two point groups\,\cite{pointg} become $C_{3v}$ and $C_3$, respectively, for a gated monolayer graphene and a biased bilayer graphene.
The loss of an inversion symmetry for a bilayer graphene under a vertical electric field has a profound effect on the formation of fractal Landau subbands in the presence of a square-scatter array potential, as can be seen 
from Eqs.\,(\ref{e8}) and (\ref{e13}) where both LL-coupling coefficients $C_n^\alpha(\xi)$ and $D_n^\alpha(\xi)$ are $\xi$ dependent and the intervalley coupling also becomes possible.   
\medskip

Compared with a monolayer graphene, a bilayer graphene can bring into additional valley mixing and site asymmetry after a perpendicular electric field ${\cal E}_0$ has been applied.
Such intervalley interference and electro-modulation effects can be seen clearly from Eq.\,(\ref{e13}) for the matrix elements of the scattering potential, i.e., 
the summation over $\xi'$ for fixed $\xi$ and changing coefficients $C_n^\alpha(\xi)$ and $D_n^\alpha(\xi)$ with $u$ and $\delta$ for $n\geq 2$.
In Figs.\,\ref{FIG:3} - \ref{FIG:5}, only a negligible electric field is employed ($\delta\sim 10^{-3}\,\hbar\omega_c$), and therefore, no visible distortions of the Hofstadter butterfly,
which results from a square 2D periodically-modulated scattering-lattice potential, can be resolved. However, as $\delta/\hbar\omega_c$ is slightly 
increased from $1\times 10^{-3}$ to $2.5\times 10^{-3}$ in Fig.\,\ref{FIG:6} for a very weak modulation with $V_0/\hbar\omega_c=0.5$ and $N=1$,
we find from Fig.\,\ref{FIG:6}$(b)$ that the previously found self-similar patterns within the third Landau band under $\delta\sim 10^{-3}\,\hbar\omega_c$ is very strongly distorted, and therefore disappears. 
\medskip

Moreover, as $\delta/\hbar\omega_c$ is further increased from $2.5\times 10^{-3}$ to $10^{-1}$ in Fig.\,\ref{FIG:7} for $V_0/\hbar\omega_c=0.5$ but $N=3$,
we find from a direct comparison between Fig.\,\ref{FIG:6}$(b)$ and Fig.\,\ref{FIG:7}$(b)$ that 
the previously observed self-repeated patterns within the fourth Landau band under $\delta\sim 10^{-3}\,\hbar\omega_c$ is destroyed completely. Meanwhile, the mixing of the third and fourth
Landau bands is seen clearly even for such a small modulation amplitude $V_0/\hbar\omega_c=0.5$ in contrast with the result in Fig.\,\ref{FIG:4}$(b)$ for $V_0/\hbar\omega_c=1.0$.

\section{Brief Summary}
\label{s4}

In conclusion, we have developed a theoretical formalism to demonstrate the Hofstadter-type fractal band structure for bilayer graphene
in the presence of a two-dimensional periodic electrostatic modulation. The current work can be viewed as a generalization of the previous reported results
based on Bloch-wave expansion approach applied to both a two-dimensional electron gas\,\cite{kuhn} and a monolayer graphene\,\cite{ourRK}.
As in previous studies\,\cite{kuhn,ourRK}, this work includes explicitly deriving a non-perturbative eigenvalue equation, finding numerical solutions which display self-repeated split Landau subbands as a function magnetic flux and periodic
subband dispersions as a function of electron wave number in a full magnetic Brillouin zone.
Both Hofstadter butterflies and band mixing effects can be displayed simultaneously for a strong scattering strength which further reduces a required magnetic field for such observations to an accessible level.
\medskip

Interestingly, we find two unique features for the bilayer-graphene system in this study.
The first one is related to a bias-modulated mixing of and an interference from two valleys (i.e., non-vanishing intervalley scattering with $\xi, \xi' = \pm 1$) in the presence of a scattering-lattice potential.
The second one, however, is associated with a lost inversion symmetry due to a perpendicular electric field, which tends to distort and even destroy the Hofstadter-type fractal band structures established by this scattering-lattice potential, 
as seen from Figs.\,\ref{FIG:6} and \ref{FIG:7}.
The dependence of Bloch-wave expansion coefficients on the applied electric field directly leads to an electro-deformation of the Hofstadter-type subband splittings, resulting in strongly distorted or even destroyed
self-repeated patterns.

\clearpage
\bibliography{BilG}


\end{document}